\documentclass{article}

\usepackage{hyperref}       % hyperlinks
\usepackage{url}            % simple URL typesetting
\usepackage{booktabs}       % professional-quality tables
\usepackage{amsfonts}       % blackboard math symbols
\usepackage{nicefrac}       % compact symbols for 1/2, etc.
\usepackage{microtype}      % microtypography
\usepackage{lipsum}		% Can be removed after putting your text content
\usepackage{graphicx}
\usepackage{natbib}
\usepackage{doi}
\usepackage{moreverb,url}
\usepackage{amsmath}
\usepackage{amssymb}
\usepackage{booktabs}
\usepackage{multirow}
\usepackage{graphicx}
\usepackage{bm}
\usepackage{xcolor}
 \usepackage{geometry}

    \newgeometry{vmargin={25mm}, hmargin={28mm,28mm}}   % set the margins

\providecommand{\keywords}[1]{\textbf{\textit{Keywords---}} #1}

\title{Analyzing differences between restricted mean survival time curves using pseudo-values}

\author{ \href{https://orcid.org/0000-0001-9358-011X}{\includegraphics[scale=0.06]{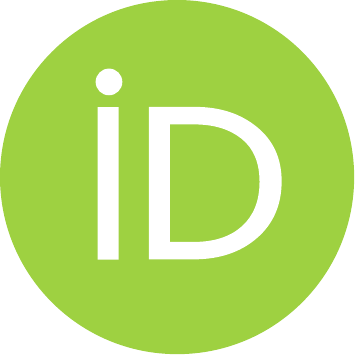}\hspace{1mm}Federico Ambrogi}\thanks{Use footnote for providing further
		information about author (webpage, alternative
		address)---\emph{not} for acknowledging funding agencies.} \\
	Department of Clinical Sciences and Community Health\\
	University of Milan\\
	Via Vanzetti 5, 20133, Milano, Italy \\
	\texttt{federico.ambrogi@unimi.it} \\
	\and
	Simona Iacobelli \\
	Department of Biology\\
	University of Rome Tor Vergata\\
	Rome, Italy\\
	\and
	\href{https://orcid.org/0000-0003-0490-0885}{\includegraphics[scale=0.06]{orcid.pdf}\hspace{1mm}Per Kragh Andersen} \\
	Department of Biostatistics \\
	University of Copenhagen\\
	\O ster Farimagsgade 5 entr. B, PO Box 2099, DK-1014 Copenhagen \\
}

\begin{document}
\maketitle

\begin{abstract}
Hazard ratios are ubiquitously used in time to event analysis to
quantify treatment effects. Although hazard ratios are invaluable for
hypothesis testing, other  measures of association, both relative and
absolute, may be used to fully elucidate study results. Restricted
mean survival time differences between groups have been advocated as
useful measures of association. Recent work focused on model-free
estimates of the difference in restricted mean survival for all
follow-up times instead of focusing on a single time horizon. In this
work a model-based alternative is proposed with estimation using
pseudo-values. A simple approach is proposed easily implementable with
available software. It is also possible to compute a confidence region
for the curve. As a by-product, the parameter 'time until treatment
equipoise' (TUTE) is also studied. Examples with crossing survival
curves will be used to illustrate the different methods together with
some simulations.
\end{abstract}


\section{Introduction}

In most clinical trials and observational studies dealing with
time-to-event as the main outcome, the measure of association used is
the hazard ratio (HR), a quantity which is typically estimated
using Cox regression.
When the proportional hazards assumption holds, Cox regression is, in fact,
the preferred method of estimation due to its efficiency.
The use of hazard ratios is well established and accepted in
biomedical literature, sometimes acritically.
In fact, many authors warned against its limitations. First of all,
its interpretation may not always be as straightforward as could be a
time based measure \citep{Spruance}. This is in part due to the
relative nature of the hazard ratio, which means that the time gained
by treated/exposed versus non-treated/non-exposed patients is not
easily evaluated as it depends also on the baseline risk. Second, if
the PH assumption is not true, reporting a single HR estimate is
obviously misleading while the reporting of an HR varying through time
does not have a simple interpretation
due to selection of patients during
follow-up \citep{Hernan1,martinussenetal}. In some situations, the proportional hazards
assumption is tenable just because of the fact that the follow-up
length is too short to show non proportionality. The need for
expressing study results in a way that people can easily understand
\citep{Greenhalghg3725} is another of the motivations that keep the
debate on the hazard ratio active.

Existing proposed alternatives include the ratio between median
survival times \citep{Spruance}, the difference of survival
probabilities at a specific time point, and the difference of the
expected survival times \citep{Uno1, R1}. The latter measure of
association is in general referred to a fixed time interval $[0,
\tau]$, i.e. the question is if there is a difference in the
restricted (at $\tau$ years) mean survival time
(RMST) \citep{klein03}. In the framework of clinical trial planning, the
comparison of RMST has interesting advantages \citep{R2}.
 % XXX quale articolo citare per l'introduzione della RMST? Il klein03 citato nel commento qui sotto?

%Fixing the comparison at a specified time interval makes in fact estimable what is called the restricted mean survival time (RMST)\citep{klein03}.

%Some authors advocated the use of RMST when the proportional hazard assumption is in fact questionable \citep{R1, R2}.

%($$CITARE Comparison of the restricted mean survival time with the hazard ratio in superiority trials with a time-to-event end point Bo Huang$$).

A possibility introduced first by Royston and Parmar \citep{R1}, is to
estimate the difference of RMST curve through the follow-up time, to
show how the treatment comparison varies in time. This approach was
developed in Zhao et al. \citep{zhao} introducing the use of
simultaneous confidence bands to make inference at all time points.

An extreme form of non PH is when the survival curves cross and in
such a situation a single measure of association, such as the hazard
ratio, is too simple to summarise the relationship between the
exposure or the treatment and the outcome.  The restricted mean
survival time (RMST) has been advocated as a possible alternative
outcome measure for such cases \citep{R1, R2}.

The case of comparison of treatments with crossing
  survival curves is particularly challenging. Noorami and
colleagues \citep{TUTE}, proposed a new measure
of impact in the framework of surgical decision making, when the
surgical decision is based on
purely prognostic considerations (and not for example aiming to relieve
symptoms). Specifically, when comparing two treatment options, such as
a surgical intervention and a conservative therapy, it is frequently
the case that the benefit of the former appears
later with respect to the benefit of the latter. The
authors proposed to assess the time elapsed after the intervention
until the expected years of life lost in the two treatment groups are
equal.
This new measure is called {\em time until treatment equipoise}
(TUTE) and it is an alternative to the, perhaps more intuitive, time
at which the two survival curves are crossing.
In general, the TUTE may be an additional useful piece of information
when using RMST curve differences in the special case of crossing
survival curves. In the work of Noorami and colleagues there are no
considerations about how to estimate TUTE and, most importantly, its
confidence interval.

In this work, we propose a simple, model-based, method to estimate the
difference in RMST curves using pseudo-values. Such a methodology
enables an easy calculation of the confidence bands for the curve and
also the possibility of adjusting for covariates. Moreover we propose an
estimate for TUTE when
the survival curves are crossing, focusing on the necessity to provide
a confidence interval for the estimate.

The use of pseudo values is not the only possibility, the flexible
parametric survival models developed by \citep{R1} or the direct regression method developed by \citep{tian}, based on weighted estimating equations,
are valid alternatives. Although software is readily
  available for all the cited approaches \citep{Klein:2008, R1, survRM2}, the approach developed here, based on
  pseudo-values, allows an easy estimation of simultaneous confidence
  bands by means of available standard software.

%for the estimation of simultaneous confidence bands.}
Some real examples are presented together with a simulation study
taking into account different scenarios.

\section{Methods}

\subsection{Restricted Mean Survival Time}
In survival analysis the time $T$ elapsed from an initial event to the
possible occurrence of a terminating event is analysed. Usually, only
a right-censored version of the random
variable $T$ is observed and, therefore, the mean value of $T$ is not
easy to estimate non-parametrically, \citep{Andersen}. As a replacement,
the $\tau$-restricted mean survival time (RMST) is defined as:

\begin{equation}
\text{RMST}(\tau) = \int_{0}^{\tau} \text{S}(t) dt \label{RM}
\end{equation}
where $\text{S}(t) = \text{P}(T>t) = \exp(- \int_{0}^{t} \lambda(u) du)$ is the survival function and $\lambda(t)$ is
the hazard function. The $\text{RMST}(\tau)$ represents the
expected lifetime, $E(T\wedge \tau)$ over a time horizon equal to $\tau$ and the
difference between $\text{RMST}(\tau)$ for different treatments
has been advocated as a useful summary measure in clinical
applications \citep{R1, R2}.  The $\text{RMST}(\tau)$ can be
estimated non-parametrically based on the Kaplan-Meier estimator or
model-based, possibly resorting to flexible regression.

%In fact, when a model for the hazard function is specified, then a model for the survival function is implied and, thereby, a model for the RMST.

In particular, the piecewise-exponential model of
  Karrison \citep{Karrison} assumes proportionality for the covariate
  effects while separate, piecewise constant, baseline hazards, are
  used for the different treatments to estimate $\text{RMST}(\tau)$. The
model was further developed by Zucker
\citep{Zucker} using Cox regression with stratification. The method is
implemented in the
function \texttt{restricted.residual.mean} in the package
\texttt{timereg}, \citep{timereg}, in the free R software,
\citep{Rs}. The function can use the Cox regression model or the Aalen
regression model to perform the calculations.
A number of different alternatives are available for model based
estimates of $\text{RMST}(\tau)$, for example one convenient possibility is
the use of flexible parametric survival models \citep{R1}. In general,
the standard errors of the RMST (or of the difference of RMST between
treatments) are obtained using the delta method, or using the
bootstrap or other resampling/simulation techniques.\\

\subsection{Direct regression models on RMST}
An alternative estimation method is to directly model RMST as a
function of covariate values. This can be achieved using the approach
of \citep{tian}, based on inverse
probability of censoring weighting, or the approach of Andersen and
colleagues \citep{and2} based on pseudo-values. Considering a binary
covariate $Z$, for example for treatment, the model with restriction time
$\tau$ can be written as:

\begin{equation}
\text{g}(\text{RMST}(\tau |Z)) = \beta Z    \label{zero}
\end{equation}
Commonly used link functions are the $log$, the $logit$ or the identity function.

We will focus on estimation based on pseudo-values as it allows to use standard software for generalized linear models.
The general theory underlying pseudo-observations was outlined in \cite{Andersen2003}.
Let $X_i$, $i = 1, \ldots, n$, be independent and identically distributed random variables,
let $\theta$ be a parameter of the form
\begin{equation}
\theta = E(f(X_i))
\end{equation}
and assume that we have an (at least approximately) unbiased estimator, $\hat{\theta}$, for this parameter. Let, furthermore, $\bm{Z}_i$, $i = 1, \ldots, n$ be independent and identically distributed covariates and define the conditional expectation
\begin{equation}
\theta = E(f(X_i) | \bm{Z}_i)
\end{equation}
The $i^{th}$ pseudo-observation is defined as
\begin{equation}
\hat{\theta}_i = n \hat{\theta} - (n-1) \widehat{\theta^{-i}}
\end{equation}
where $\widehat{\theta^{-i}}$ is \lq\lq the leave-one-out\rq\rq  estimator for $\theta$ based on $X_j$, $j \neq i$. If all $X_i$
are observed then $\theta$ may be estimated by the average of the $f(X_i)$ in which case $\hat{\theta}_i$ is
simply $f(X_i)$. We will use this approach when only a censored
sample of the $X_i$ is available.
A regression model for the parameter $\theta$ corresponds to a specification of how $\theta_i$
depends on $Z_i$ and this may done via a generalized linear model
\begin{equation}
g({\theta}_i) = \bm{\beta}^{T} \bm{Z}_i
\end{equation}
where the matrix $\bm{Z}$ contains a column of 1, corresponding to the intercept. The regression coefficients $\bm{\beta}$ can be estimated using generalized estimating equations
\begin{equation}
U(\bm{\beta}) = \sum_{i=1} ^n U_{i}(\bm{\beta}) = \sum_{i=1} ^n \left( \frac{\partial}{\partial \bm{\beta}} g^{-1} (\bm{\beta}^T \bm{Z}_{i}) \bm{V}_i ^{-1} (\widehat{\bm{\theta}}_{(i)} -  g^{-1} (\bm{\beta}^T \bm{Z}_{i})) \right)     \label{RSE}
\end{equation}
In the general situation $\theta$ may be multivariate and $\bm{V}_i$ is the working covariance matrix.
\cite{Andersen2003} argued that the variances of $\bm{\beta}$ can be obtained by the standard sandwich estimator
\begin{equation}
\hat{\sum} = I(\hat{\bm{ \beta}})^{-1} \hat{var} \{U(\bm{\beta}) \} I(\hat{\bm{ \beta}})^{-1}
\end{equation}
where
\begin{equation}
 I(\bm{ \beta}) =  \sum_{i=1} ^n  \left( \frac{\partial g^{-1}(\bm{\beta}^T \bm{Z}_{i})}{\partial \bm{\beta}} \right)^{T} \bm{V}_i ^{-1}  \left( \frac{\partial g^{-1} (\bm{\beta}^T \bm{Z}_{i})}{\partial \bm{\beta}}  \right)
\end{equation}
\begin{equation}
 \hat{var} \{U(\bm{\beta}) \} =  \sum_{i=1} ^n U_{i}(\bm{\beta}) U_{i}(\bm{\beta})^{T}
\end{equation}
After the computation of pseudo-observations, parameter estimates and their standard errors can be computed using standard statistical software for generalized estimating equations, though the standard errors may be slightly conservative.
In fact, \cite{overgaard} presented a general asymptotic theory of estimates from estimating functions based on pseudo-observations demonstrating, under some regularity conditions, consistency and asymptotic normality of the estimates. The ordinary sandwich estimator is however not consistent, leading to an overestimate of the standard errors. The demonstration is derived for real-valued pseudo-observations, but, as the Authors say, it can be generalized to handle vector-valued pseudo-observations.

For the restricted mean we have $\theta = E(X \wedge \tau) = \int_0 ^{\tau} S(t) dt$, and we use the estimator obtained by plugging in the Kaplan-Meier estimator \citep{and2}. For this estimator, results stated in \citep{overgaard} are valid under the assumption of censoring independent of event times and covariates.
The $i^{th}$ pseudo-value at time $\tau$ is therefore defined as:
\begin{equation}
\widehat{\theta}_{(\tau \, i)} = n  \int_{0}^{\tau} \widehat{\text{S}}(t) dt - (n-1)
\int_{0}^{\tau} \widehat{\text{S}}^{-i}(t) dt
\end{equation}
where $\widehat{\text{S}}^{-i}(t)$ is the Kaplan-Meier estimator excluding subject $i$.

Instead of considering a single $\tau$ as in \cite{and2}, we consider a finite grid of $M$ time points $\tau_{1}, \ldots, \tau_{j}, \ldots,
\tau_{M}$ and we compute the pseudo-values for the
$i^{th}$ subject at each $\tau_{j}$.
Time points can be selected as quantiles of the event time
distribution, e.g. $M$ could be chosen to have approximately $10$
events for each pseudo-value while, in general, it is not useful to
have more than $15$-$20$ time points.
A regression model for a vector valued $\hat{\bm{\theta}}_i$, with components calculated at several $\tau$-values, must include terms for time and possibly an interaction term
between covariates and time to account for possible time-varying covariate effects. This was already done for model based on pseudo-values with applications to competing risks, for example in \cite{KAcomp}.
The model can be written as:
\begin{equation}
 \text{g}(\hat{\bm{\theta}}_i) = \text{g}(\text{RMST}(\pmb{\tau} |\bm{Z})) = \text{h}(\pmb{\tau}) + \bm{\beta} \bm{Z} + \bm{\gamma} \bm{Z}
\text{f}(\pmb{\tau}).  \label{one}
\end{equation}
The interaction term has the purpose of modelling non time-constant effects
on the scale of the link function used. For example, using the $log$ link, when the ratio of restricted mean survival times is constant through the follow-up time, the interaction term can be excluded.
It is worth pointing out that, when considering the identity link, a
constant difference in the restricted means through time is not
plausible. The difference will always start at zero and then
eventually change. The same considerations apply for the inclusion
of covariates in this model. When looking for adjusted estimates,
covariates should be included together with their
time-dependent effects, i.e. interactions with time. 

For the RMST it is
customary to use the identity link function ($g(u)=u$) or,
alternatively, the "log" link \citep{LOGAN}. In the following the
identity link will be used and the regression model will therefore be
specified as:

\begin{equation}
\text{RMST}(\tau_j\mid Z_i) = \theta_{(\tau_{j} \, i)} = \alpha_{0} + \alpha_{j} \text{I}_j + \beta Z_{i} + \gamma_{j} \text{I}_j Z_{i}
\end{equation}
where the $\text{I}_{j}, j=2,\dots,M$ are $M-1$ indicator functions for
estimation of the baseline function $\text{h}(\tau_j)$. The same indicator functions are used to model time-varying covariate effects.

Considering the estimating equations (\ref{RSE}), $\bm{V}_i$ is the working variance-covariance matrix, which can be conveniently set to the identity matrix. In the application presented in \cite{and2} both $\widehat{\theta}_{(i)}$  and  $V_i$
were scalar and the model did not include the effect of time as in model (\ref{zero}).

The estimate of the difference in RMST through follow-up between treatments is given by the step function

\begin{equation}
\text{R}(\tau_j) = \Delta(\tau_j | Z) =  \beta + \gamma_{j} \text{I}_{j}
\end{equation}
changing value at each time selected for the computation of the pseudo-values.

The variance of $\text{R}(\tau_j)$ can be estimated for each $\tau_{j}$
from model results as in standard GEE modelling (though, as already said, this may be
slightly conservative \citep{overgaard}).
For example, for time $\tau_{j}$, the variance of interest, can be
computed using an $M$-dimensional basis vector, with $1$ in position
$j$ and the variance-covariance matrix for the coefficients $\beta$
and $\gamma_{j}$:

\[
{\small
\begingroup % keep the change local
\setlength\arraycolsep{1pt}
\begin{pmatrix}
0 & \ldots & 1 & \ldots  & 0\\
\end{pmatrix}
\begin{pmatrix}
\text{V}(\widehat{\beta}) & \ldots  & \text{Cov}(\widehat{\beta}, \widehat{\gamma}_{j}) & \ldots&\\
\ldots & \ldots & \ldots & \ldots & \\
\ldots & \text{Cov}(\widehat{\gamma}_{j-1}, \widehat{\gamma}_{j}) & \text{V}(\widehat{\gamma}_{j}) &\ldots & \\
\ldots & \ldots & \ldots  & \ldots &  \\
\ldots & \ldots & \ldots & \text{V}(\widehat{\gamma}_{M-1}) & \\
\end{pmatrix}
\begin{pmatrix}
0 \\
\vdots \\
1 \\
\vdots \\
0
\end{pmatrix}
\endgroup
}
\]
where $\text{V}(\cdot)$ stands for the cluster-robust variance, while
$\text{Cov}(\cdot, \cdot)$ stands  for the cluster-robust covariance of two
random variables. The cluster is given by each subject
  repeated at multiple time points.

The function $\text{R}(\tau_j)$ can also be obtained by incorporating a
smooth spline basis into the
regression model. In this case, without loss of generality,
considering just two basis functions (for ease of notation)
$\text{B}_{1}(t)$ and $\text{B}_{2}(t)$ to model the RMST through time,
$h(t)$, the regression model can be written as:

\begin{align}
\text{RMST}(t \mid Z_i) & = \alpha_{0} + \alpha_{1} \text{B}_{1}(t) + \alpha_{2}
\text{B}_{2}(t) + \beta Z_{i} + \nonumber \\
& \qquad \gamma_{1} \text{B}_{1}(t) Z_{i} +  \gamma_{2} \text{B}_{2}(t) Z_{i} \label{eq1}
\end{align}
and the difference in RMST curves is given by the smooth function:

\begin{equation}
\text{R}(t) = \Delta(t | Z) =  \beta + \gamma_{1} \text{B}_{1}(t) +  \gamma_{2} \text{B}_{2}(t)
\end{equation}
The cluster robust variance at time $t$ of the estimate $\widehat{\text{R}}(t)$,
$V(\widehat{\text{R}}(t))$ can be computed as:
%\[\arraycolsep=1.4pt\def\arraystretch{2.2}
\[
{\small
\begingroup % keep the change local
\setlength\arraycolsep{1pt}
\begin{pmatrix}
1 & \text{B}_{1}(t) & \text{B}_{2}(t)\\
\end{pmatrix}
\begin{pmatrix}
\text{V}(\widehat{\beta}) & \text{Cov}(\widehat{\beta}, \widehat{\gamma}_{1}) & \text{Cov}(\widehat{\beta}, \widehat{\gamma}_{2})\\
\text{Cov}(\widehat{\beta}, \widehat{\gamma}_{1}) & \text{V}(\widehat{\gamma}_{1}) & \text{Cov}(\widehat{\gamma}_{1}, \widehat{\gamma}_{2})\\
\text{Cov}(\widehat{\beta}, \widehat{\gamma}_{2}) & \text{Cov}(\widehat{\gamma}_{1}, \widehat{\gamma}_{2}) & \text{V}(\widehat{\gamma}_{2})
 \end{pmatrix}
 \begin{pmatrix}
1 \\
\text{B}_{1}(t) \\
\text{B}_{2}(t)
\end{pmatrix}
\endgroup
}
\]
To compute a pointwise confidence interval for this function it is
possible to resort to percentile bootstrap or to the asymptotic
normality of the model estimates.
As remarked in \citep{AndePerme}, similar arguments can be applied in
the case of RMST, provided a limited set of time points is used for
the analysis.

Based on the asymptotic normality of the model estimates, the
asymptotic pointwise $95\%$ confidence interval of $\text{R}(t)$ is given by
$[\text{R}_{lo}(t);
\text{R}_{up}(t)] = \widehat{\beta} + \widehat{\gamma}_{1} \text{B}_{1}(t) +
\widehat{\gamma}_{2} \text{B}_{2}(t) \pm 1.96 \sqrt{\text{V}(\widehat{\text{R}}(t))}$.

In addition, again relying on the asymptotic normality of the model
estimates, it is possible to adopt the approach developed in
\citep{SIGPM} on simultaneous inference in general parametric models to
estimate the simultaneous $95\%$ confidence interval of $\text{R}(t)$,
i.e. the confidence region for the curve.
In order to ensure a coverage probability of at least $95\%$ for the
entire curve, an appropriate critical value $u_{95\%}$ must be chosen
instead of the $1.96$. The value can be chosen such that $\text{P}(t_{max}
\leqslant u_{95\%}) = 95\%$, where:

\begin{equation}
 t_{max} = \text{sup}_{t\in[a,b]}\frac{[\widehat{\beta} +
   \widehat{\gamma}_{1} \text{B}_{1}(t) + \widehat{\gamma}_{2} \text{B}_{2}(t) ] -
   \text{R}(t)}{\sqrt{\text{V}(\widehat{\text{R}}(t))}}
  \end{equation}
where the limits of the interval $[a, b]$ span the follow-up time of
interest or, more strictly, corresponds to the minimum and maximum
times used to compute pseudo-values.
In order to compute the $u_{95\%}$ value, the supremum of the function
can be obtained using an equally spaced grid of time points $[a
\leqslant t_{1} \leqslant t_{1} \leqslant \cdots \leqslant t_{k} =
b]$.
The obtained value should be sufficiently close to the true value and
this approach makes it possible to use standard software for the
calculation \citep{multiple}.

One important aspect for the implementation of the
method is the choice of the spline function. Standard
  B-spline bases
can have weird effects outside boundary knots. To
obtain more stable estimates of the difference between RMST along the whole
follow-up, it seems to be useful to
use some form of restriction, such as those implemented in natural
splines (see for example \texttt{ns} function in \texttt{R}).

\subsection{TUTE}

The time, $\tau^{*}$, at which (if ever) the $\text{RMST}(\tau)$ for two
treatment groups are equal was called the time until treatment
equipoise (TUTE) by Noorami and colleagues \citep{TUTE}.
Specifically, if $Z$ is the covariate for the
treatment (e.g., $0$ conservative therapy, $1$ surgical therapy), we are
interested in the quantity

\begin{equation}
\tau^{*} = \inf_{\tau>0} \int_{0}^{\tau} (\text{S}(t|Z=1) - \text{S}(t|Z=0)) dt = 0.
\end{equation}
When the survival curves are crossing, the $R(t)$ function starts
at $0$, then decreases (increases), reaches a minimum (maximum), and
then starts increasing (decreasing). The root of this function
provides an estimate for TUTE, $\tau^{*} = \text{R}^{-1}(0)$.

The time $\tau{*}$ can be calculated using numerical iterative
techniques such as those
implemented in numerical root finder \texttt{uniroot} function in
R. When there is no crossing, $\tau^{*} = \infty$.

To compute the point-wise confidence interval for the TUTE one simple
possibility would be to resort to the bootstrap. For each bootstrap
sample, $b$, the TUTE is calculated, $\tau^{*}_{b}$ and then a
confidence
interval can be calculated using the percentile method.
When performing the bootstrap, there could be samples showing no
crossing between the curves. In these samples the TUTE is
$\infty$. Ignoring these samples would result in an underestimation of
the limits of the confidence interval of TUTE. If the
  estimated TUTE equals $\infty$ in more than $5\%$ of the cases, there is no
evidence for the existence of a finite TUTE.

An alternative approach relies on the asymptotic normality of the
model estimates. In this case, the $95\%$ confidence interval for TUTE
can be obtained by searching for the zeroes of the upper and lower
point-wise confidence intervals of $\text{R}(t)$, i.e. $[\text{R}_{up}^{-1}(0);
\text{R}_{lo}^{-1}(0)]$.

When the upper limit never crosses the $x$ axis, the 95\% confidence
interval is one-sided open to the right. It is also possible that the
lower limit never crosses the $x$ axis, in this case the lower limit
is $0$.
%The same is valid for the confidence region.

\section{Model Complexity}
\label{s:CV}

Considering model (\ref{eq1}), only $2$ spline bases were used for convenience. However, the choice of the amount of smoothing, i.e. the complexity of the spline, to model the
baseline RMST, is an open problem.  One empirical solution is to use the
quasi likelihood for the model with pseudo-values: 

\begin{equation}
  \text{QL} = \sum_{i=1}^{N} \sum_{j=1}^{M}  (\mathbf{ \widehat{ \theta}}_{(\tau_{j} i)} -  \text{RMST}(\tau_j\mid Z_i) )^{2}.
\end{equation}

This is in line with the use of pseudo-residuals for the evaluation of the goodness of fit used in \citep{Checking}.  More principled approaches are emerging in literature, \cite{GoF}, and will hopefully improve also the possibilities of model selection.

To select the number of knots the quasi information criterion (QIC) can then be adopted \citep{pan}:
\begin{equation}
\text{QIC} = -2 \text{QL} + 2  \; \mbox{tr}(\text{R}^{-1} \text{V})
\end{equation}
where $\mathbf{N}$ is the na\"{\i}ve variance estimate while
$\mathbf{V}$ is the cluster robust variance estimate.
This is the approach used in the example and the simulations. However,
as the selection does not regard the working correlation structure, the
trace could simply be replaced by twice the number of model
parameters.\\

\section{Results}

\subsection{Simulation}

\subsubsection{Use of multiple restriction times}

The use of a vector of pseudo-observations at a grid of $M$ time points is standard practice in applications of pseudo observations involving multi state models. In applications with RMST only a single time point, i.e. the restriction time, is used.
In the applications presented here we are using a vector of restriction times, and therefore multiple pseudo-observations per subject, in order to estimate the difference between RMST curves through follow-up time together with a confidence band.
In this simulation we want to investigate the behaviour of the model estimated with multiple pseudo-observations per subject by comparing it to the standard pseudo value model with just a single restriction time and with the approach proposed by \cite{tian} based on weighted equations. In particular, we use the same simulation design proposed in \cite{and2}.

Weibull distributed life times were generated with scale parameter $\lambda_{i} = \exp{(\beta_b Z_i)}$ and shape parameter $\delta = 0.5$, $1$ or $2$. Here, $Z_i$ is binary with $pr(Z_i = 1)= 0.5$ and
$\beta_b = 0$ or $1$. Exponential censoring at $25$\% was superimposed and the
restricted mean life time at $\tau$ was estimated for values of $\tau$ at the $p^{th}$ percentile when $\beta_b = 0$, i.e. $\tau = (-log(1-p))^{1/\delta}$ for $p = 0.75$ and $0.9$. The true value of the restricted mean is
\begin{equation}
\int_{0}^{\tau} \exp(- \lambda t^{\delta}) dt = \frac{1}{\delta} \lambda^{-1/\delta} \left[ \Gamma(\frac{1}{\delta}, 0) - \Gamma(\frac{1}{\delta}, \lambda (-log(1-p)))  \right]
\end{equation}
where $\Gamma(a, x)$ is the incomplete gamma function.
The baseline RMST is therefore,

$\beta_{0} = \frac{1}{\delta} \left[ \Gamma(\frac{1}{\delta}, 0) - \Gamma(\frac{1}{\delta}, (-log(1-p)))  \right]$,
while the Z effect, i.e. the difference in RMST between $Z=1$ and $Z=0$ is given by $\beta_{1}=\frac{1}{\delta} \exp^{-1/\delta} \left[ \Gamma(\frac{1}{\delta}, 0) - \Gamma(\frac{1}{\delta}, \exp(1) (-log(1-p)))  \right] - \beta_{0}$.

For the standard model with pseudo-values proposed by \cite{and2} and for the model of \cite{tian}, $\beta_{0}$ and $\beta_{1}$ correspond to the intercept and to the coefficient of $Z$.
For the model with a vector of pseudo values, $16$ times were selected at quantiles of the failure time distribution, starting from the minimum until the $99^{th}$ percentile, and the pseudo-observations for each subject were calculated. Natural splines were used to estimate the baseline RMST and an interaction between splines bases and $Z$ was used to estimate the curve $\text{R}(t)$. The value of baseline RMST and of $\text{R}(t)$ at time $\tau$ is then calculated.

Each combination was replicated 1000 times. Simulations in which the last simulated event time was less than $\tau$ were excluded. This happened in an important number of times with setting $\delta=0.5$ and $\beta_{b}=1$ ($72$ times with $P=0.75$ and $472$ times with $P=0.90$ when $N=250$ and $325$ times with $P=0.90$ when $N=1000$). Also with setting $\delta=1$ and $\beta_{b}=1$ this happened $140$ times with $P=0.90$ when $N=250$ and $95$ times with $P=0.90$ when $N=1000$. In these two settings it happened also that the last restriction time of the model estimated using a vector of pseudo-values (the last restriction time is placed at the 99\% percentile of the failure time distribution) was less than $\tau$ ($\delta=0.5$ and $\beta_{b}=1$: $22$ times with $P=0.75$ and $988$ times with $P=0.90$ when $N=250$ and $1000$ times with $P=0.90$ when $N=1000$; $\delta=1$ and $\beta_{b}=1$ this happened $279$ times with $P=0.90$ when $N=250$ and $38$ times with $P=0.90$ when $N=1000$). Results are shown in table \ref{firsts}. The biases are everywhere quite small for all the methods compared, with the exception of the model with the vector pseudo-values in setting $\delta=0.5$ and $\beta_{b}=1$, especially for the $90^{th}$ percentile. This is due to the fact that for the direct model with a vector of pseudo-observations the estimates at $\tau$ are obtained in extrapolation. This is easy to avoid in applications.

Another important point regards complexity selection. The number of spline bases was chosen in each simulated data with $QIC$ in a range between $3$ and $12$. However, results are not changing fixing the the degrees of freedom to $3$ in each simulation (not shown).

%The empirical standard deviations of the estimates were in close agreement with the standard errors based on (11) (not shown).

\begin{table*}[ht]
\caption{Comparison of the regression model estimated with pseudo-values using a single restriction time at $\tau$ (PV scalar) or multiple restriction times at quantiles of failure time distribution (PV vector), with the approach of \citep{tian}.
For PV vector $16$ pseudo values are used in each setting. QIC was used
  to select the degrees of freedom of the splines (from a minimum of
  $3$ to a maximum of $12$). Two different sample size are considered
  ($250$ and $1000$ with 25\% censoring).}
\begin{tabular}{@{}lllc|llllll|@{}}
\cmidrule(l){5-10}
                                               &                                                 &       & \multicolumn{1}{l|}{} & \multicolumn{3}{c|}{$P$=0.75}                                                               & \multicolumn{3}{c|}{$P$=0.90}                                                               \\ \cmidrule(l){3-10}
                                               & \multicolumn{1}{l|}{}                           & $\delta$ & $\beta_{b}$                  & \multicolumn{1}{c}{PV scalar} & \multicolumn{1}{c}{Tian} & \multicolumn{1}{c|}{PV vector} & \multicolumn{1}{c}{PV scalar} & \multicolumn{1}{c}{Tian} & \multicolumn{1}{c|}{PV vector} \\ \midrule
\multicolumn{1}{|l|}{\multirow{12}{*}{N=250}}  & \multicolumn{1}{l|}{\multirow{6}{*}{Baseline}} & 0.5   & 0                     & -0.001                        & -0.001                   & \multicolumn{1}{l|}{-0.001}    & 0.003                         & 0.003                    & 0.026                          \\
\multicolumn{1}{|l|}{}                         & \multicolumn{1}{l|}{}                           & 0.5   & 1                     & 0.000                         & 0.001                    & \multicolumn{1}{l|}{0.014}     & 0.026                         & 0.025                    & 0.137                          \\
\multicolumn{1}{|l|}{}                         & \multicolumn{1}{l|}{}                           & 1     & 0                     & 0.000                         & 0.000                    & \multicolumn{1}{l|}{-0.001}    & 0.000                         & 0.000                    & 0.011                          \\
\multicolumn{1}{|l|}{}                         & \multicolumn{1}{l|}{}                           & 1     & 1                     & -0.003                        & -0.003                   & \multicolumn{1}{l|}{0.002}     & -0.005                        & -0.005                   & 0.004                          \\
\multicolumn{1}{|l|}{}                         & \multicolumn{1}{l|}{}                           & 2     & 0                     & -0.001                        & -0.001                   & \multicolumn{1}{l|}{-0.001}    & -0.001                        & -0.001                   & -0.001                         \\
\multicolumn{1}{|l|}{}                         & \multicolumn{1}{l|}{}                           & 2     & 1                     & -0.001                        & -0.001                   & \multicolumn{1}{l|}{0.001}     & -0.001                        & -0.001                   & 0.001                          \\ \cmidrule(l){2-10}
\multicolumn{1}{|l|}{}                         & \multicolumn{1}{l|}{\multirow{6}{*}{$Z$ effect}}     & 0.5   & 0                     & 0.001                         & 0.001                    & \multicolumn{1}{l|}{0.001}     & 0.003                         & 0.003                    & 0.003                          \\
\multicolumn{1}{|l|}{}                         & \multicolumn{1}{l|}{}                           & 0.5   & 1                     & 0.000                         & -0.001                   & \multicolumn{1}{l|}{-0.006}    & 0.043                         & 0.036                    & 0.181                          \\
\multicolumn{1}{|l|}{}                         & \multicolumn{1}{l|}{}                           & 1     & 0                     & 0.000                         & 0.000                    & \multicolumn{1}{l|}{0.000}     & 0.000                         & 0.000                    & 0.000                          \\
\multicolumn{1}{|l|}{}                         & \multicolumn{1}{l|}{}                           & 1     & 1                     & -0.003                        & -0.003                   & \multicolumn{1}{l|}{-0.013}    & -0.004                        & -0.005                   & -0.002                         \\
\multicolumn{1}{|l|}{}                         & \multicolumn{1}{l|}{}                           & 2     & 0                     & 0.002                         & 0.002                    & \multicolumn{1}{l|}{-0.002}    & 0.002                         & 0.002                    & 0.002                          \\
\multicolumn{1}{|l|}{}                         & \multicolumn{1}{l|}{}                           & 2     & 1                     & 0.002                         & 0.002                    & \multicolumn{1}{l|}{-0.002}    & 0.002                         & 0.002                    & 0.002                          \\ \midrule
\multicolumn{1}{|l|}{\multirow{12}{*}{N=1000}} & \multicolumn{1}{l|}{\multirow{6}{*}{Baseline}} & 0.5   & 0                     & 0.005                         & 0.005                    & \multicolumn{1}{l|}{0.002}     & -0.003                        & -0.003                   & 0.023                          \\
\multicolumn{1}{|l|}{}                         & \multicolumn{1}{l|}{}                           & 0.5   & 1                     & 0.001                         & 0.001                    & \multicolumn{1}{l|}{0.014}     & 0.001                         & 0.001                    & 0.084                          \\
\multicolumn{1}{|l|}{}                         & \multicolumn{1}{l|}{}                           & 1     & 0                     & 0.0003                        & 0.0003                   & \multicolumn{1}{l|}{-0.002}    & 0.0001                        & 0.0001                   & 0.013                          \\
\multicolumn{1}{|l|}{}                         & \multicolumn{1}{l|}{}                           & 1     & 1                     & -0.001                        & -0.0005                  & \multicolumn{1}{l|}{0.005}     & -0.001                        & -0.001                   & 0.002                          \\
\multicolumn{1}{|l|}{}                         & \multicolumn{1}{l|}{}                           & 2     & 0                     & 0.0003                        & 0.0003                   & \multicolumn{1}{l|}{0.0002}    & 0.0003                        & 0.0003                   & 0.0002                         \\
\multicolumn{1}{|l|}{}                         & \multicolumn{1}{l|}{}                           & 2     & 1                     & 0.0003                        & 0.0003                   & \multicolumn{1}{l|}{0.003}     & 0.0003                        & 0.0003                   & 0.003                          \\ \cmidrule(l){2-10}
\multicolumn{1}{|l|}{}                         & \multicolumn{1}{l|}{\multirow{6}{*}{$Z$ effect}}     & 0.5   & 0                     & 0.001                         & 0.001                    & \multicolumn{1}{l|}{0.001}     & 0.005                         & 0.005                    & 0.005                          \\
\multicolumn{1}{|l|}{}                         & \multicolumn{1}{l|}{}                           & 0.5   & 1                     & 0.002                         & 0.001                    & \multicolumn{1}{l|}{-0.007}    & 0.005                         & 0.004                    & 0.120                          \\
\multicolumn{1}{|l|}{}                         & \multicolumn{1}{l|}{}                           & 1     & 0                     & 0.001                         & 0.001                    & \multicolumn{1}{l|}{0.001}     & 0.0002                        & 0.0002                   & 0.0005                         \\
\multicolumn{1}{|l|}{}                         & \multicolumn{1}{l|}{}                           & 1     & 1                     & -0.0002                       & -0.0002                  & \multicolumn{1}{l|}{-0.011}    & -0.001                        & -0.001                   & -0.009                         \\
\multicolumn{1}{|l|}{}                         & \multicolumn{1}{l|}{}                           & 2     & 0                     & 0.0002                        & 0.0002                   & \multicolumn{1}{l|}{-0.0002}   & 0.0002                        & 0.0002                   & 0.0002                         \\
\multicolumn{1}{|l|}{}                         & \multicolumn{1}{l|}{}                           & 2     & 1                     & 0.0004                        & 0.0004                   & \multicolumn{1}{l|}{-0.0003}   & 0.0004                        & 0.0004                   & 0.0003                         \\ \bottomrule
\end{tabular}
\label{firsts}
\end{table*}

\subsubsection{RMST curve and TUTE}

In order to examine the proposed method to estimate the RMST
difference curve and the corresponding TUTE, if any, different
simulations were performed.
The simulated survival functions are represented in Figure
(\ref{fig:1}) while the details of the simulations are reported in
the appendix.
The first scenario regards a typical situation in which the
proportional hazards assumption is not reasonable and the curves are
crossing at the end of the considered follow-up.
The scenarios from $2$ to $4$ are taken from \citep{pone1}
where different testing procedures were compared in the presence of
crossing survival curves. The crossing is at different probability levels.
%The other scenario comes from supposing a time dependent treatment
%effect directly causing the crossing of the curves. This last
%scenario corresponds to the specification of regression model
%(\ref{one}) with an interaction between treatment and time.
The fifth simulation scenario is used to mimic the crossing of
survival curves found in the third clinical application. In this
scenario, the two survival curves are practically superimposed at the
beginning, then separate and then cross.

\begin{figure*}
\centering
\includegraphics[scale=.58]{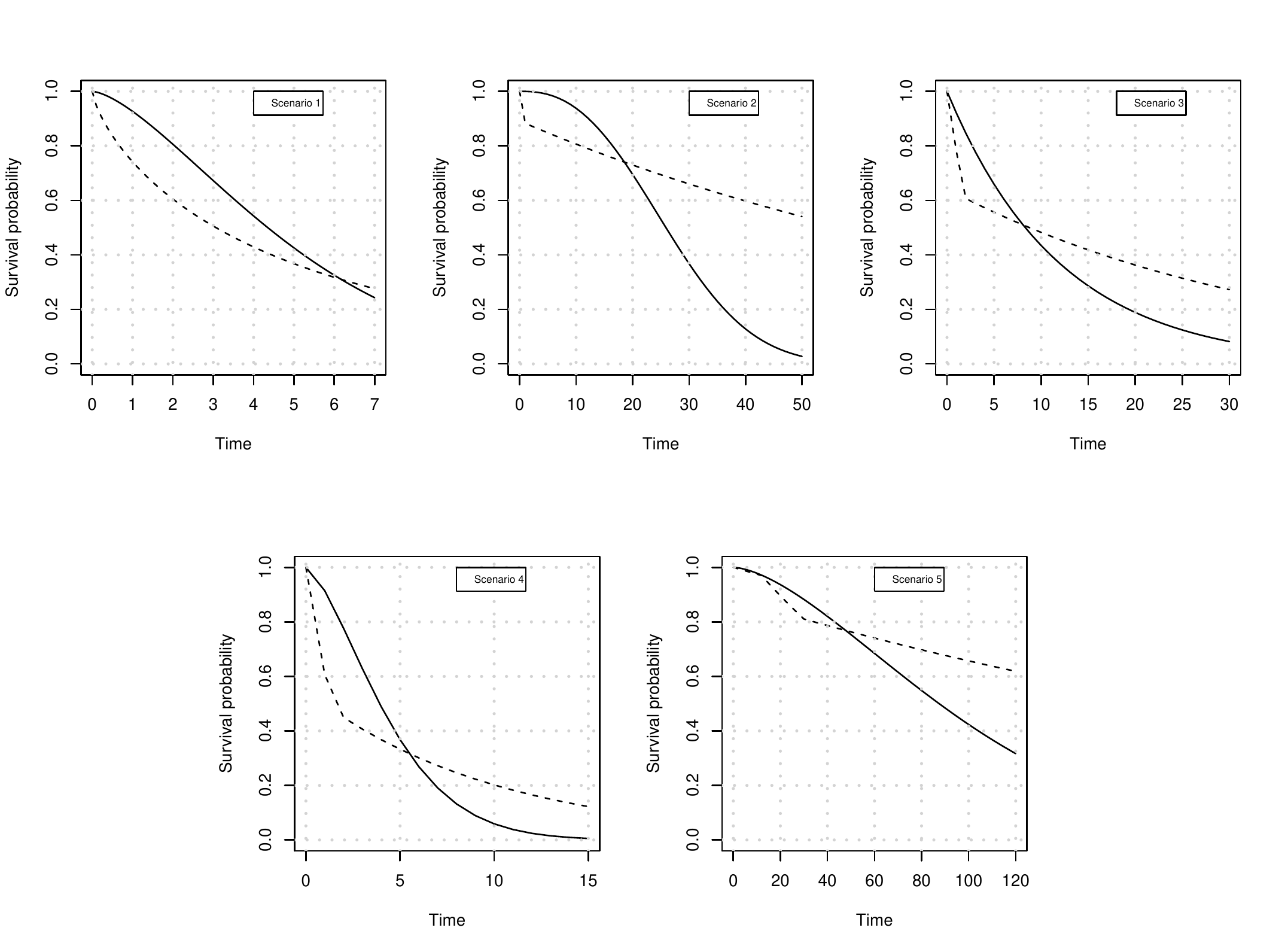}
\caption{solid and dotted lines represent the survival curves used in the 4 different scenarios. The detailed description is in the appendix. }
\label{fig:1}
\end{figure*}

For each simulation the RMST difference curve was calculated, using
the KM estimator, together with its 95\% confidence band, according to
the method of \citep{zhao}. The curve and the 95\% confidence band were
also estimated using pseudo-values and GEE regression.
To calculate the confidence band with pseudo-values a grid of $10$ or
more, equally spaced, time points was used. In general the result is
quite stable using $10$ or more time points. The function
\texttt{glht} from the package \texttt{multcomp} (\citep{SIGPM}) was
used to compute the confidence band.
To evaluate the coverage, it was checked if the band included the true
RMST difference value, for all the time points of the grid.
The average length of the band was also computed together with the
average bias in the estimate. Results are reported in Table
(\ref{tab1}).
For the pseudo-value model, $16$ time points were considered, at
quantiles of the event time distribution. Simulations with a
different number of time points did not show different results (not
shown). However, using a small number of time points for the analysis
prevents the possibility to increase the spline complexity for the
estimation of the function $\text{R}(t)$. \\

For each scenario, natural splines with different degrees of freedom
were used varying from $4$ to $12$.
QIC was used to select the degrees of freedom in each simulated data
set. Boundary knots were set to the minimum and maximum time used for
pseudo-value calculation.
%The function \texttt{glht} from the package \texttt{multcomp}
%(\citep{SIGPM}) was used to compute the confidence band using the same
%fine grid of time points used to check the non parametric band. When
%the function returned an error a model with one less degree of
%freedom was tried. If also the model with $4$ degrees of freedom
%failed the calculation was aborted. \\

Results from simulations show that the regression model with
pseudo-values has results comparable with those of the non-parametric
estimators. The coverage of the band is good and approximates quite
closely the desired 95\%.

In simulation scenarios $2$, $3$ and $5$, the TUTE was also
calculated. Results are given in terms of bias with respect to the
true TUTE and in terms of root mean squared error (RMSE), as the
length of the confidence interval is not always finite. Also in this
case, the model estimated using pseudo-values gives comparable results
with that of the non-parametric estimator. The confidence interval was
also calculated using the 95\% point-wise confidence interval of the
RMST difference curve. The coverage is approximately the one
expected. The average length of the 95\% confidence interval is not
reported as in some simulations, the interval is open to the right.
In fact, for a sample size of $200$ per group, the 95\% confidence
interval of TUTE was open to the right in $3\%$, $23\%$ and $28\%$ of
the simulations for scenarios $2$, $3$ and $5$ respectively. These
percentages became $0\%$, $2\%$ and $4\%$ with $400$ subjects per
group.
In simulation scenario $4$ the TUTE is at time $14.6$,
  i.e. when two survival curves are already below the $20\%$
  probability. This is due to the fact that the curves cross at a low
  probability level (about $30\%$). In such cases the TUTE is not
  interesting from a clinical viewpoint and the estimation was not
  carried out.

\begin{table*}[ht]
\caption{Comparison of the results obtained through the non-parametric
  estimator with that from the regression model on pseudo
  values. Sixteen pseudo values are used in each setting. QIC was used
  to select the degrees of freedom of the splines (from a minimum of
  $4$ to a maximum of $12$). The different scenarios are described in
  figure (\ref{fig:1}). Two different sample size are considered
  ($200$ and $400$ per group with 20\% censoring).}
\centering
{\scriptsize
\begin{tabular}{c@{\hskip .05in}c|c@{\hskip .05in}c@{\hskip
      .05in}c|c@{\hskip .05in}c@{\hskip .05in}c|c@{\hskip
      .05in}c@{\hskip .05in}c|c@{\hskip .05in}c@{\hskip .05in}c}

\toprule
                     &                      & \multicolumn{6}{c}{Non
                       Parametric}                                 &
                     \multicolumn{6}{c}{Pseudo-Values}
                     \\
\midrule
\multicolumn{1}{}{} & \multicolumn{1}{l}{} & \multicolumn{3}{c}{Curve} & \multicolumn{3}{c}{TUTE} & \multicolumn{3}{c}{Curve} & \multicolumn{3}{c}{TUTE }  \\
\midrule
                     &  Scenario           &  Bias  & Coverage &
                     Length  &  Bias  &  Coverage  &  RMSE & Bias  &
                     Coverage &  Length  &  Bias  &  Coverage &  RMSE
                     \\
\midrule
\multirow{5}{*}{\rotatebox{90}{200}}
  &1 & 0.095 & 0.941 & 0.643 &           &          &           & 0.096 & 0.941 & 0.612 &           &         &  \\
  &2 & 0.719 & 0.931 & 4.297 & 0.346 &  0.950 & 10.5 & 0.717 & 0.943 & 4.425 & 0.247 & 0.949 & 10.5  \\
  &3 & 0.702 & 0.937 & 4.634 & -0.302 & 0.948 & 13.9 & 0.704 & 0.952 & 4.524 & -0.441 & 0.945 & 13.8  \\
  &4 & 0.172 & 0.928 & 1.140 &           &           &           & 0.172 & 0.938 & 1.115 &           &           &  \\
  &5 & 1.434 & 0.973 & 9.886 & 0.444 & 1.000 & 262.7 & 1.446 & 0.954 & 9.499 & 0.019 & 0.978 & 254.9\\
    \midrule
\multirow{5}{*}{\rotatebox{90}{400}}
  &1 & 0.069 & 0.946 & 0.464 &           &       &           & 0.069 & 0.928 & 0.437 &           &       &  \\
  &2 & 0.502 & 0.939 & 3.138 & 0.038 & 0.952 & 5.2 & 0.503 & 0.937 & 3.225 & 0.018 & 0.952 & 5.3\\
  &3 & 0.502 & 0.948 & 3.539 & 0.141 & 0.952 & 8.7 & 0.506 & 0.929 & 3.398 & 0.061 & 0.952 & 8.6 \\
  &4 & 0.125 & 0.944 & 0.866 &           &       &           & 0.125 & 0.945 & 0.839 &           &       &  \\
  &5 & 1.033 & 0.965 & 7.143 & 0.044 & 1.000 & 183.9 & 1.038 & 0.942 & 6.931 & -0.171 & 0.976 & 180.3\\
    \hline
\end{tabular}
\label{tab1}
}
\end{table*}

\subsection{Applications}

\subsubsection{The CSL1 Trial in Liver Cirrhosis}

\begin{figure*}[ht]
\centering
\includegraphics[scale=.58]{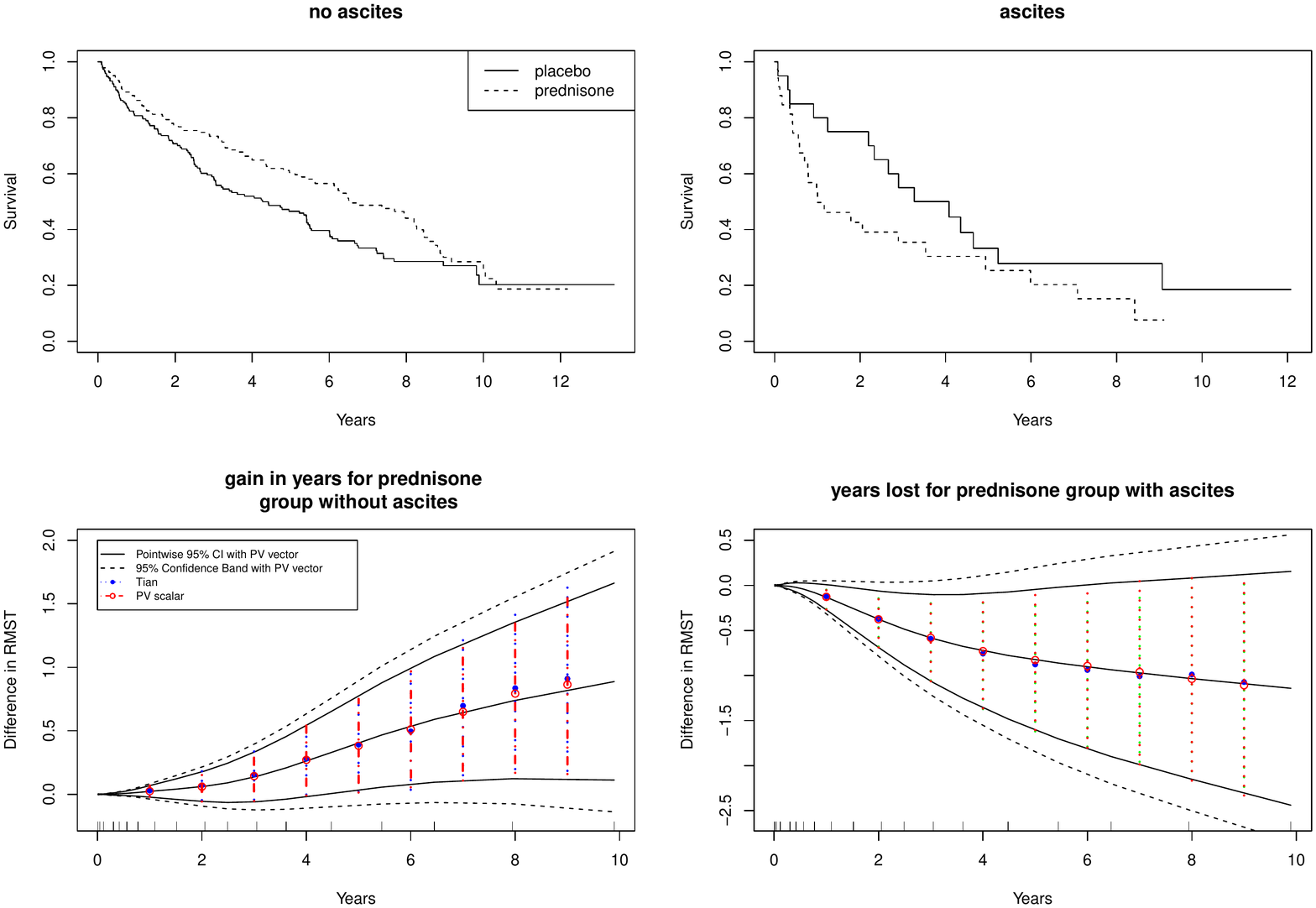}
\caption{CSL1 Trial in Liver Cirrhosis. On the top the Kaplan-Meier survival curves for the ascites and no ascites groups of patients.
On the bottom the difference of RMST in the two groups for different restriction times and direct modelling approaches. The small
    vertical lines on the $x$ axis represent the times used to calculate the
    pseudo-values.}
\label{figCSL}
\end{figure*}

The CSL1 trial was already analysed in \cite{and2} with pseudo-observations considering both mean and restricted mean survival time, with restriction at $5$ years. The randomized trial studied the effect of prednisone on survival in patients with liver cirrhosis \citep{Christensen:1985}.
An interesting finding was that only patients without ascites seemed to benefit from the treatment. The reanalysis presented here aims to compare three different approaches to the analysis of restricted mean: the method based on pseudo values and the weighted regression of Tian with restricted mean at a specified $\tau$ (in years) and the method with pseudo values with multiple restriction times used to estimate the RMST curve.
According to the last method, $16$ pseudo-times, at quantiles of the failure time distribution, were used to calculate the pseudo-observations for each patient. Then a regression model with identity link function and interaction between ascites and treatment was estimated. The model was also adjusted by age and all effects were time dependent as required by the identity link. Five degrees of freedom were used for modelling the baseline RMST as suggested by the QIC criterion. Also according to QIC the model with interaction between ascites and treatment has to be preferred with respect to the model without ($19633$ vs $19883$).
In figure \ref{figCSL} it is reported, together with the Kaplan-Meier curves in the ascites and no ascites groups, the curve estimated with multiple pseudo values with its $95\%$ point-wise confidence interval and band. Moreover it is reported in red the result obtained applying a regression model based on pseudo-values considering a single restriction time, starting from $1$ year, then $2$ years, until $9$ years. The same procedure was applied for the model with weights using the \texttt{R} function \texttt{rmst2} from the package \texttt{survRM2} \citep{survRM2}.
Results from the three different modelling approaches are quite similar with the advantage for the first method of making possible the estimation of the simultaneous confidence band for the curve of all restriction times.

\subsubsection{Gefitinib or Carboplatin-Paclitaxel in Pulmonary Adenocarcinoma.}

The second application regards a randomised controlled clinical trial
comparing Gefitinib with Carboplatin-Paclitaxel in Pulmonary
Adenocarcinoma \citep{pmid19692680}. As the data are not freely
available, reconstructed data were used, according to the procedure
outlined in \citep{RKM} starting from the results of the paper. In
particular, panel A of Figure 2 of the article, depicting Kaplan-Meier
curves for progression-free survival for the overall population, was
reproduced.  The aim of the study was to demonstrate non-inferiority
of Gefitinib compared to Carboplatin-Paclitaxel. The patients were
randomized to Gefitinib (n=$609$) or to Carboplatin plus Paclitaxel
(n=608).

\begin{figure*}
\centering
\includegraphics[scale=.58]{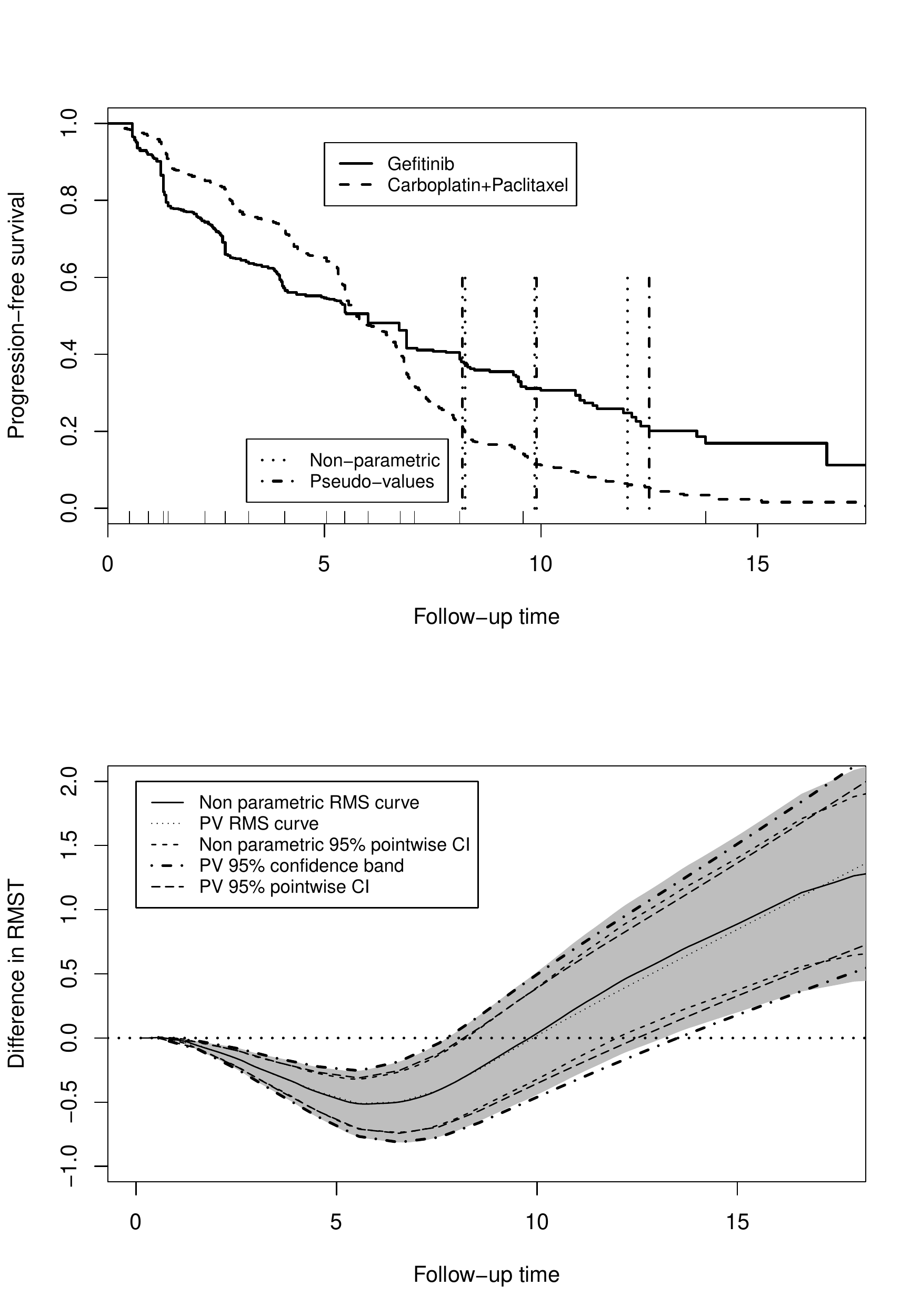}
\caption{Adenocarcinoma:
the top panels show the progression free survival
    curves with the TUTE estimated non-parametrically (dot) and with
    pseudo-values (dotdash). Also the corresponding 95\% confidence
    intervals are shown for the adenocarcinoma application. The small vertical
    lines on the $x$ axis represent the times used to calculate the
    pseudo-values. The bottom panels show the difference in RMST
    between the two compared groups estimated non-parametrically (continuous) and with the
  pseudo values (dot). The grey area corresponds to the 95\% confidence band
  estimated with the non-parametric method. The thick dotdashed lines
  are the 95\% confidence bands estimated with pseudo-values.
 The figure shows also the point-wise lower and upper $95\%$
 confidence interval for the non-parametric method (dashed) and for
 the pseudo-values method (longdashed).}
\label{fig2}
\end{figure*}

The reconstructed data are reported in the top panel of Figure
(\ref{fig2}). Kaplan-Meier curves cross at about $5.7$ months. The
RMST curve is reported in the bottom panel of Figure
(\ref{fig2}). The curve shows the difference in the area under the
progression-free survival curve between Gefitinib and
Carboplatin-Paclitaxel groups. The difference is negative at the
beginning becoming positive afterwards by crossing the x axis. Both
the upper and the lower limit of the confidence band show the same
pattern. In this situation it is possible to compute the 95\%
confidence interval for the TUTE. The estimated TUTE using the
non-parametric method is $9.85$ months with 95\% confidence interval
($8.19-12.10$).

Only in 8 bootstrap samples ($1.6\%$) the survival curves were without
crossing. The mean TUTE of the bootstrap replicates is $9.86$ months
with 95\% confidence interval equal to ($8.13-12.82$). The estimated
TUTE with the pseudo-value regression model is $9.90$ months
($8.19-12.50$), very similar to the non parametric and bootstrap
estimates.
 Boundary knots of the natural splines of the pseudo-value regression
 model were placed at $0.26$ and $13.80$ months. Twelve degrees of
 freedom were used for the natural spline according to QIC. For the
 calculation of the confidence band a grid starting at $1$ until $17$
 months, spaced by $1$ was used.

In the original application, the HR was used to
  demonstrate the difference between the two treatments. This is
  clearly inappropriate. The use of the RMST difference returns a more
  complete picture of the advantage of Gefitinib vs
  Carboplatin-Paclitaxel. The time to equipoise is equal to
  approximately $10$ months (with a closed 95\% confidence interval)
  with a follow-up length that hardly is greater than $15$ months. In
  fact the crossing of the curves occurs at low probability levels and
  the TUTE is quite high. In practice, although there is an advantage
  of Gefitinib its clinical importance could be questioned. These
  considerations are made using simulated and not the actual
  data. Moreover to make a complete clinical consideration the age of
  the patients should be taken into account.

\subsubsection{EBMT-NMAM2000 study.}

The third example refers to the NMAM2000 trial comparing tandem
autologous/reduced intensity conditioning allogeneic transplantation
(auto+allo) to autologous transplantation alone (auto) on an
intent-to-treat basis. The analysis and the corresponding clinical
considerations are published in \citep{iacobelli} while those presented
here are illustrative considerations for the statistical methods
presented.

\begin{figure*}
\centering
\includegraphics[scale=.58]{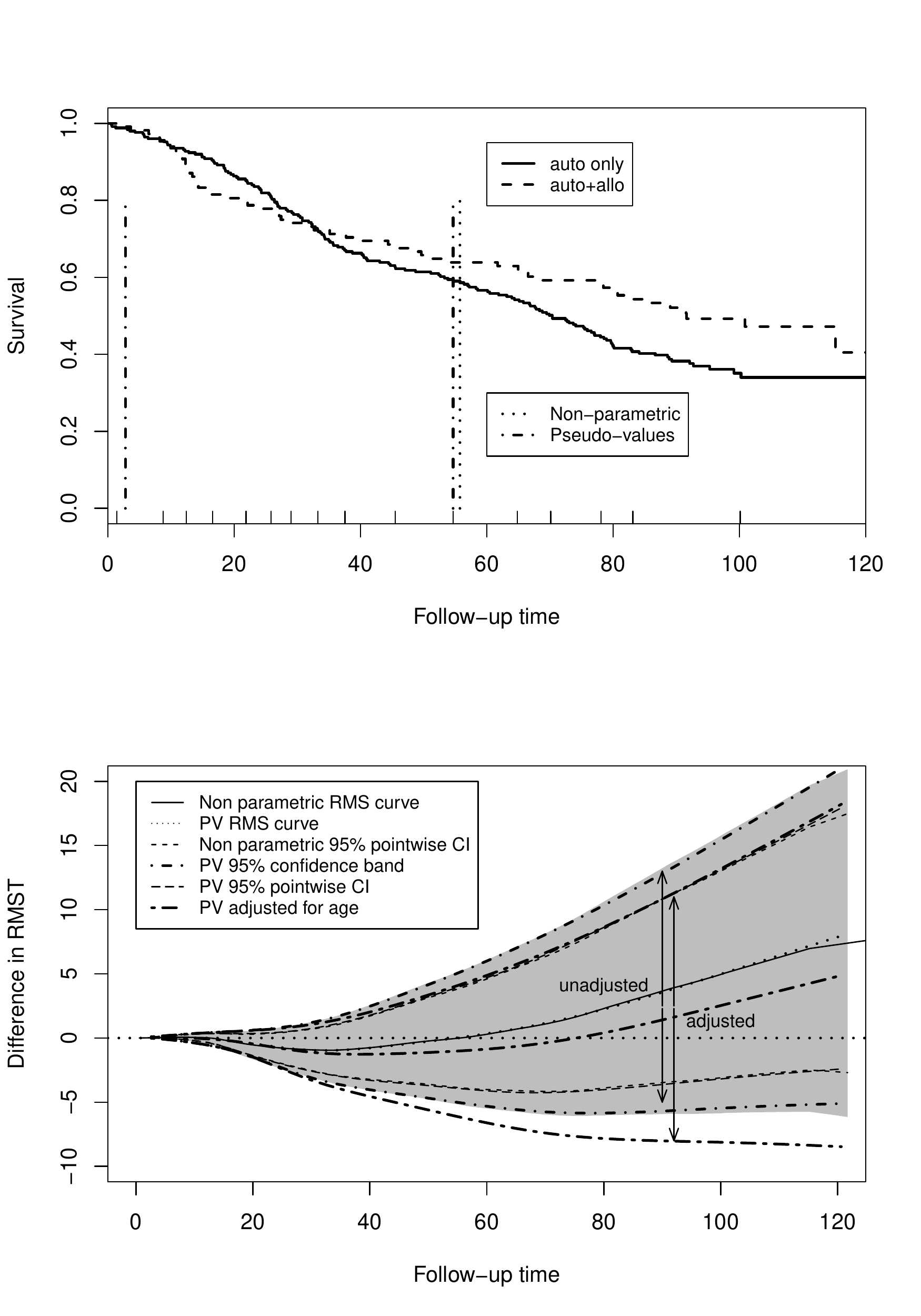}
\caption{Multiple Myeloma:
The top panels show the overall survival
    curves with the TUTE estimated non-parametrically (dot) and with
    pseudo-values (dotdash). Also the corresponding 95\% confidence
    intervals are shown for the adenocarcinoma application. The small
    lines on the $x$ axis represent the times used to calculate the
    pseudo-values. The bottom panels show the difference in RMST
    between the
  two compared groups estimated non-parametrically (continuous) and with the
  pseudo values (dot). The grey area corresponds to the 95\% confidence band
  estimated with the non-parametric method. The thick dotdashed lines
  are the 95\% confidence bands estimated with pseudo-values.
 The figure shows also the point-wise lower and upper $95\%$
 confidence interval for the non-parametric method (dashed) and for
 the pseudo-values method (longdashed). In the bottom panel the difference in
 RMST adjusted for age is also reported (twodash). }
\label{fig3}
\end{figure*}

The overall survival probability curves are reported in the top
panel of Figure (\ref{fig3}).
The curves have a similar pattern for the first year, then they
separate with auto+allo group having more events than auto group,
later the curves cross at about 33 months where auto+allo seems
superior to the auto group.

The RMST curve is reported in the bottom panel of Figure
(\ref{fig3}). The curve shows the difference in the area under the
overall survival curve between the auto-allo and auto groups. The
difference is near $0$ at the beginning, then becomes negative turning
definitely positive afterwards by crossing the x axis. However, the
upper and the lower limits of the confidence band never cross the x
axis. In this situation it is only possible to compute a one sided
95\% confidence interval for the TUTE.
The estimated TUTE using the non parametric method is $55.74$ months
with an open to the right confidence interval ($0-\infty$). The
estimated TUTE with the pseudo-value regression model is $54.83$
months ($2.77-\infty$), in accordance with the non parametric
estimates.
In this case, in $86\%$ of the bootstrap samples the survival curves were without crossing. The mean TUTE of the bootstrap replicates for which there is a crossing is $22.02$ months, showing a clear underestimation.
Boundary knots of the natural splines of the pseudo-value regression
model were placed at $0.56$ and $100.12$ months. Five degrees of
freedom were used for the natural spline according to QIC. For the
calculation of the confidence band a grid starting at $3$ until $120$
months, spaced by $2$ was used.

In terms of difference in RMST between the two compared treatments, it
is quite clear that the time for equipoise is reached relatively
early, before $5$ years of follow-up. However, in this application the
sample size is not sufficient to appreciate the difference between the
two treatments. In fact, there are $249$ patients in the auto only
group and only $108$ in the auto+allo group. According to the
simulations, probably at least twice the sample size would serve to
observe a significant pattern.

A regression model adjusted for age was also fitted. Age was
categorized using $55$ as cutoff. Considering the identity link, age
was inserted into the model together with its time dependent
effect. Again $5$ degrees of freedom were used for the natural spline
of time. The association between RMST and age (time-dependent) was
significant (p=0.033, df=6). The adjusted difference between RMST is
reported in Figure (\ref{fig3}). It is possible to see how
the adjusted curve crosses the y axis later in time with respect to
the unadjusted estimate.

\clearpage

\section{Discussion}

The use of HR in clinical studies is generally accepted as a useful
measure of association. Notwithstanding this, the debate about the use
of HR is always active, especially because its use is strictly tied
with the Cox regression model and the assumption of proportional
hazards. In fact, especially with long follow-up length, the
tenability of this assumption becomes more questionable
\citep{R2}. Moreover, concerns about the clinical usefulness of the HR
are always present, as it is difficult to translate an HR in terms of
clinical benefit.
In general, as no single measure can be useful in all circumstances,
it is advisable not to simply rely on the HR to quantify the
association in time to event analysis.

Many alternatives have been proposed in the literature. For example,
the difference (or ratio) in survival probabilities at a specific time
point, or the difference (or ratio) of RMST at a specific time point
could be taken into consideration. These proposals have the obvious
drawback that a single time point should be selected for the
analysis. In some circumstances, as in the application presented on
multiple myeloma, a clinically relevant time horizon is present but
this is not the case in many clinical studies.

In this perspective, the proposal to look at how the
difference of RMST varies through follow-up is
particularly appealing and dates back, at least, to the work of
Royston and Parmar \citep{R1}. The main caveat when looking at the
entire curve is that it would be appropriate to resort to a confidence
band instead of the point-wise confidence limit. This was the object
of a recent proposal from Zhao and colleagues \citep{zhao}, based on
the Kaplan-Meier estimator, that can be defined model-free according
to Uno and colleagues \citep{Uno1}.

In this work a simple model-based method, relying on pseudo-values,
was proposed to provide inference on the RMST
difference curve based on a confidence band. The method is in good agreement with the estimates obtained by direct regression models fixing one restriction time. Moreover, the
method is flexible enough to reproduce the results of the model-free
method when no covariates are considered. The proposed methodology allows
to adjust for covariates. This could be particularly
  important as demonstrated in the NMAM2000 study where the estimated
  curve changes when adjusting for age, as expected.

In principle, other flexible regression models could be used for the
same purpose. In practice, the estimation based of pseudo-values is
very easy to be implemented and can rely completely on standard
available software, also for the confidence band calculation. One
drawback is that it is necessary to choose how many time points to use
for pseudo-values calculations and how to space them. Although this
aspect should be further investigated, it seems that varying the
number of time points does not alter substantially the results.

One important clinical situation for which the use of RMST has been
frequently advocated is in presence of
crossing survival curves. This is a long debated subject. Crossing survival
curves indicate a gross violation of the proportional hazards
assumption, where the Cox model cannot be conveniently used. In this
situation a summary statement using the hazard ratio  can be
misleading as the factor under study has an implicit time-dependent pattern. One
possibility is to resort to statistical test procedures to answer the
question of which treatment option produces the better long term
results. In this case some strategies have been proposed, see for
example \citep{LOGAN2}. A second interesting suggestion relates to the
difference in RMST
at a pre-specified time point, as an effective
summary measure of effect along a
  time interval \citep{R2}.

Further, instead of looking at the difference at one
  specific time, the use of the entire curve, together with its
confidence band, can be useful for understanding the pattern of
survival time lost or gained during the
follow-up. In presence of crossing survival curves it is also
interesting to study the time at which the RMST difference curve
crosses the x axis. In fact, the so-called time until treatment
equipoise is the time point, during follow-up, at which the RMST of
the two treatment options under comparison are equal. TUTE was
introduced by Noorami and colleagues \citep{TUTE} as an aid in decision
making for surgery of asymptomatic patients.

In fact, TUTE can be a useful complement for the
  description of the treatment effect when survival curves are
  crossing, for example when comparing an aggressive treatment to a
  more conservative one.  If the survival curves do not cross (no
finite TUTE), then clearly one treatment is superior to the other
(provided the difference is statistically significant). In the case of
crossing curves, a very long TUTE is an indicator
  that the divergence after the crossing is not sufficiently large to
  balance the initial disadvantage of the more aggressive strategy. Of
  notice, if the crossing occurs at low probability levels (such as in
  simulation scenario 4) there is no point in computing the
  TUTE. Conversely, a short TUTE, close to the crossing time,
  corresponds in general to survival curves that diverge markedly
  after the crossover point.

  Besides being a summary indicator of the type of
  crossing curves, TUTE is informative for decision making. Both TUTE
  and the crossover time point can be interpreted as the maximum time
  horizon that a patient shall look at in order to opt for the
  conservative treatment instead of the more aggressive treatment, the
  difference being in the type of utility function considered. For
  TUTE, the utility is measured as life expectancy, for the crossover
  time the utility is measured as chance of long-term survival. For an
  elderly  patient, or in case of a disease with dismal outcome, the
  average life duration might be more relevant than the probability of
  survival. In this situation, a patient willing to optimize his/her
  utility along a time span longer than TUTE should choose the more
  aggressive strategy. As a final remark for a correct interpretation,
  neither TUTE nor the crossover time indicate a time from which the
  outcome is better for the patients who receive the aggressive
  treatment than with for those treated with the conservative one,
  because both are measured from the beginning of the follow-up. For a
  dynamic comparison of the two strategies, proper approaches could be
  based  on hazard functions, or on updated predictions, such as
  landmark curves.

From a methodological viewpoint, here we stressed the importance of
providing a confidence interval for TUTE
and we compared two estimation methods for TUTE (and its 95\%
confidence interval):
\begin{itemize}
\item The first method relies on the use of an estimate of the
  survival function. The method can be non-parametric or
  model-based. After obtaining $\widehat{S}(t)$, it is possible to
  estimate the difference between RMST as a function of time and
  finding the root of such a function.
\item The second method relies on the direct modelling of RMST through
  a regression model using pseudo-values with a time-varying
  coefficient, $\beta(t)$. The
  root of the function $\beta(t)$ is the required TUTE.
\end{itemize}
The second method has the potential advantage of including different
covariates, thus proving an adjusted measure of
TUTE. This is particularly important
for example since age should
probably be considered when making
  considerations involving TUTE.
Regarding the first method, when multiple covariates are under study,
one possibility is to fix one specific covariate pattern (such as
identifying a low risk or high risk patient), and computing the TUTE
for such a pattern. Another possibility would be to use an averaging
method, such as the average corrected group prognosis method,
\citep{pmid11572743} (equal to the '$g$-formula' \citep{HernanRobins})
to evaluate the difference between two treatment options for
example. These considerations about covariates are valid, also, for
the curve estimate of the difference of RMST between groups through
follow-up time. The investigation of the different issues arising when
producing adjusted estimates of effect was not done in the present
paper.

\section{Acknowledgement}
We gratefully acknowledge the European Society for Blood and Marrow
Transplantation (EBMT) for making available data of NMAM2000 trial.

The work was partially supported by Italian Ministry of Education, Universities and Research project PRIN 2017, prot. 20178S4EK9\_004, Innovative Statistical methods in biomedical research on biomarkers: from their identification to their use in clinical practice.

\section{Appendix}
\subsection{Simulations details}
Event times were simulated according to the following specifications:

%$\lambda(t) = \left\{ \begin{array}{c} \lambda_{1}  \,\,\,\,\, t<t_{1}  \\  \lambda_{2}   \,\,\,\,\, t \geq t_{2}  \end{array} \right .$
%The integrated hazard is
%$H(t) = \left\{ \begin{array}{c} \lambda_{1} t  \,\,\,\,\, t<t_{1}  \\ \lambda_{1} t_{1} + \lambda_{2} (t-t_{1})   \,\,\,\,\, t \geq t_{2}  \end{array} \right .$
%The inverse of the cumulative hazard is therefore
%$H^{-1}(x) = \left\{ \begin{array}{c} \frac{x}{  \lambda_{1}}   \,\,\,\,\, x<\lambda_{1} t_{1}  \\ t_{1} + \frac{x-\lambda_{1} t_{1}}{\lambda_{2}}   \,\,\,\,\, x \geq \lambda_{1} t_{1}  \end{array} \right .$
%where $x = -\log(U)$, and $U \sim U[0,1]$, see \citep{AUSTINsim}.

For the simulations scenarios see \citep{pone1}:
\begin{itemize}
\item Scenario 1: (1) Weibull with parameters ($0.18; 1.5$) and (2) Weibull with parameters ($0.20; 0.75$).
In this scenario the crossing is toward the end of the considered follow-up and TUTE is of no interest.
\item  Scenario 2: (1) Weibull with parameters ($2.5; 30$) and (2) piecewise exponential with $\lambda=0.125 \, I(t<1) + 0.01 \, I(t\geqslant 1)$.
In this scenario the crossing of the curves is at time $18.55$ while the TUTE is $30.93$.
\item  Scenario 3:(1) exponential with $\lambda=1/12$ and a piecewise exponential with $\lambda=0.25 \, I(t<2) + \frac{1}{35} \, I(t\geqslant 2)$.
In this scenario the crossing of the curves is at time $8.09$ while the TUTE is $17.75$.
\item  Scenario 4:(1) Weibull with parameters ($1.5; 5$) and (2) piecewise exponential with $\lambda=0.5 \, I(t<1.5) + 0.1 \, I(t\geqslant 1.5)$.
In this scenario the crossing of the curves is at time $5.48$ while the TUTE is $14.57$.
\item  Scenario 5:(1) Weibull with parameters ($1.6; 110$) and (2) piecewise exponential with $\lambda=0.0025 \, I(t<12) + 0.01 \, I(12 \leqslant t <30) + 0.003 \, I(t\geqslant 30)$. This scenario is similar to the data from the EBMT-NMAM2000 study and the TUTE is at $73$.
\end{itemize}

For all simulations a 20\% random uniform censoring was considered.

%\subsection{Reproducible research}
%Source R code to perform the computations is available as Supporting Information
%on the journalÕs web page. The code uses the publicly available dataset \texttt{pbc}.

%\bibliographystyle{SageV.bst}
%
\bibliographystyle{unsrtnat}
\bibliography{TUTE281020.bib}

\end{document}